\documentclass[lettersize, journal]{IEEEtran}
\usepackage{amsmath,amsfonts,amssymb}
\usepackage{algorithmic}
\usepackage{algorithm}
\usepackage{array}
\usepackage{textcomp}
\usepackage{url}
\usepackage[colorlinks]{hyperref}
\usepackage{cite}
\usepackage{stfloats}  
\usepackage{subfig}
\usepackage{graphicx}
\newcommand{\figref}[1]{Fig.~\ref{#1}}
\usepackage{verbatim}  
\usepackage[utf8]{inputenc}

\begin{document}

\title{Fluid Antenna-Empowered Receive \\Spatial Modulation}

\author{
Xinghao Guo, Yin Xu,~\IEEEmembership{Senior Member,~IEEE,} Dazhi He,~\IEEEmembership{Senior Member,~IEEE,} Cixiao Zhang, \\Hanjiang Hong,~\IEEEmembership{Member,~IEEE,} Kai-Kit Wong,~\IEEEmembership{Fellow,~IEEE,} Chan-Byoung Chae,~\IEEEmembership{Fellow,~IEEE,} \\Wenjun Zhang,~\IEEEmembership{Fellow,~IEEE,} and Yiyan Wu,~\IEEEmembership{Life Fellow,~IEEE}

\thanks{
    This work was supported in part by the National Key Research and Development Project of China under Grant 2023YFF0904603; in part by the National Natural Science Foundation of China Program under Grant 62422111, Grant 62371291, and Grant 62271316; in part by IITP/NRF Grant RS-2024-00428780 and Grant 2022R1A5A1027646. \textit{(Corresponding author: Yin Xu.)}

    X. Guo, Y. Xu, C. Zhang, D. He, and W. Zhang are with the Cooperative Medianet Innovation Center, Shanghai Jiao Tong University, Shanghai, 200240 China. D. He is also affiliated with Pengcheng Laboratory, Shenzhen, 518055 China (e-mail: {guoxinghao, xuyin, hedazhi, cixiaozhang, zhangwenjun}@sjtu.edu.cn).

    H. Hong and K.-K. Wong are with the Department of Electronic and Electrical Engineering, University College London, Torrington Place, WC1E7JE, United Kingdom. K.-K. Wong is also affiliated with Yonsei Frontier Lab, Yonsei University, Seoul, 03722 Korea (e-mail: \{hanjiang.hong, kai-kit.wong\}@ucl.ac.uk). 

    C.-B. Chae is with the School of Integrated Technology, Yonsei University, Seoul, 03722 South Korea (e-mail: cbchae@yonsei.ac.kr).
    
    Y. Wu is with the Department of Electrical and Computer Engineering, Western University, London, ON N6A 3K7 Canada (e-mail: yiyan.wu@ieee.org).
}

}

\markboth{Journal of \LaTeX\ Class Files,~Vol.~14, No.~8, August~20xx}%
{Shell \MakeLowercase{\textit{et al.}}: A Sample Article Using IEEEtran.cls for IEEE Journals}


\maketitle

\begin{abstract}
Fluid antenna (FA), as an emerging antenna technology, fully exploits spatial diversity. This paper integrates FA with the receive spatial modulation (RSM) scheme and proposes a novel FA-empowered RSM (FA-RSM) system. In this system, the transmitter is equipped with an FA that simultaneously activates multiple ports to transmit precoded signals. 
We address three key challenges in the FA-RSM system: port selection, theoretical analysis, and detection. 
First, for port selection, an optimal algorithm from a capacity maximization perspective are proposed, followed by two low-complexity alternatives. Second, for theoretical analysis, performance evaluation metrics are provided for port selection, which demonstrate that increasing the number of activated ports enhances system performance. Third, regarding detection, two low-complexity detectors are proposed. 
Simulation results confirm that the FA-RSM system significantly outperforms the conventional RSM system. The proposed low-complexity port selection algorithms facilitate minimal performance degradation. 
Moreover, while activating additional ports improves performance, the gain gradually saturates due to inherent spatial correlation, highlighting the importance of effective port selection in reducing system complexity and cost. Finally, both proposed detectors achieve near-optimal detection performance with low computational complexity, emphasizing the receiver-friendly nature of the FA-RSM system.
\end{abstract}


\section{Introduction} \label{Sec-Intro}

\IEEEPARstart{F}{uture} communication systems are expected to achieve higher spectral efficiency (SE) and enhanced stability. A wide range of promising technologies, including advanced coding and modulation \cite{ref-Hong-NUC}, multiple-input multiple-output (MIMO) \cite{ref-MIMO, ref-Liu-MIMO}, non-orthogonal multiple access (NOMA) \cite{ref-Ou-NOMA}, are extensively studied to meet the demands of wireless communication. 
Among these, MIMO technology has been widely applied in numerous communication systems, owing to its high throughput enabled by deploying multiple antennas at both the transmitter and receiver. However, traditional MIMO systems deploy fixed-position antennas (FPA), which limits their ability to fully exploit the spatial variations of the wireless channel within the area covered by the antennas.

Recently, an emerging technology, fluid antenna system (FAS), has been proposed to address this issue \cite{ref-FAS, ref-FAS-Survey, ref-Hong-FAMA}. FAS refers to systems that can dynamically adjust the shape or position of antennas through software. 
Practical implementations, similar to the FAS mechanism, have already been realized using flexible conductive materials, stepper motors or reconfigurable pixels \cite{ref-LiquidAnt,ref-MotorAnt,ref-ReconPixel}. 
The original fluid antenna (FA) consists of a single radio frequency (RF) chain and multiple preset positions, also known as ports, which are distributed within the given area \cite{ref-FAS-Analysis}. Since the ports are typically densely distributed, spatial correlation plays a significant role in FAS \cite{ref-FAS-BlkChan}. By performing port selection based on signal observations, FAS can switch to the better port to achieve a higher data rate and lower interference, thereby effectively exploiting the spatial diversity \cite{ref-FAS-PS}. 
FAS has inspired research on related antenna technologies, such as movable antenna systems and flexible antenna arrays \cite{ref-MA, ref-FAA}, all of which can be referred to as next-generation reconfigurable antenna (NGRA) systems.

Due to the significant gains demonstrated by FAS, numerous studies have explored its integration with other emerging technologies, such as backscatter communication (BC) \cite{ref-FAS-BC}, physical layer security (PLS) \cite{ref-FAS-PLS}, and reconfigurable intelligent surface (RIS) \cite{ref-FAS-RIS}.
Notably, \cite{ref-MIMO-FAS} applies FAS to MIMO systems and develops the MIMO-FAS, where multiple ports are simultaneously activated to enhance performance, and a suboptimal algorithm for port selection, beamforming, and power allocation is proposed. \cite{ref-MIMO-FAS-SCSI} proposes an algorithm for precoding design and port selection in MIMO-FAS using statistical channel state information (CSI). \cite{ref-MIMO-FAS-JCR} introduces a new joint convex relaxation (JCR) problem for MIMO-FAS and designs two corresponding optimization algorithms. \cite{ref-MIMO-FAS-Heuristic} employs physics-inspired heuristics to address the port selection problem in MIMO-FAS. 

On the other hand, spatial modulation (SM), as an innovative form of MIMO technology, utilizes the indices of transmit antennas to convey additional information \cite{ref-SM}. Compared to traditional MIMO systems, SM systems offer higher energy efficiency, lower complexity, and reduced costs. To improve the SE of SM, generalized SM (GSM) has been proposed, which relaxes the constraint on the number of transmit antennas \cite{ref-GSM}. SM and GSM are transmitter-friendly schemes suitable for uplink transmission. 
In contrast, precoding-aided SM (PSM), also known as receive SM (RSM), employs classical zero-forcing (ZF) precoding or minimum mean square error (MMSE) precoding to convey additional information via the indices of receive antennas \cite{ref-RSM, ref-RSM-ICSI}. RSM is receiver-friendly, making it more suitable for downlink transmission. Similarly, generalized RSM (GRSM) has been proposed to enhance the SE of RSM and relax the constraint on the number of receive antennas \cite{ref-GRSM}. 

Based on the principles of RSM, most research assumes that the number of transmit antennas is greater than or equal to the number of receive antennas.  Therefore, for over-determined MIMO systems (where the number of transmit antennas is fewer than the number of receive antennas), a series of receive antenna selection (RAS) algorithms have been proposed to facilitate the application of RSM in such systems. Specifically, \cite{ref-RSM-FRAS} proposes a fast incremental RAS algorithm based on the greedy approach. To achieve a better trade-off between performance and complexity, \cite{ref-RSM-ERAS} proposes two efficient RAS algorithms. \cite{ref-RSM-RAS-LR} proposes a lattice reduction (LR)-based ZF precoder for the RSM system and designs the corresponding LR-based RAS algorithm. 
Additionally, \cite{ref-RSM-TAS} and \cite{ref-RSM-ETAS} propose several transmit antenna selection (TAS) algorithms for the RSM system and investigate the impact of TAS on transmission performance. 
The results in \cite{ref-RSM-TAS} and \cite{ref-RSM-ETAS} indicate that, while reducing the number of activated transmit antennas lowers hardware costs, it inevitably leads to significant performance degradation. This makes the study of TAS in traditional RSM systems less meaningful.
However, all studies on antenna selection for RSM systems do not consider the spatial correlation among antennas, which could yield different results. Moreover, these studies focus on RSM systems that employ ZF precoding, which may cause reduced effectiveness when other precoding techniques, such as the commonly used MMSE precoding, are applied.

More recently, the mechanism of SM has been applied to FAS. 
Initially, \cite{ref-IM-FA-Conf, ref-IM-FA} proposes an FA-enabled index modulation (FA-IM) system, where FA is deployed at the transmitter, and the index of the activated port is used to transmit additional information. 
\cite{ref-FA-IM-NN} employs neural network to achieve fast classification of index patterns for the FA-IM system. \cite{ref-FA-PIM} designs a port pre-selection scheme for the FA-IM system. 
\cite{ref-FA-IM} applies FA-IM to MIMO systems, where multiple ports are simultaneously activated based on the principles of GSM, significantly enhancing SE. Subsequently, \cite{My-FAG-IM} improves the system in \cite{ref-FA-IM} by considering the spatial correlation of FAS, thereby enhancing its robustness. 
Furthermore, \cite{ref-FA-IM-RIS} applies FA-IM to RIS-assisted millimeter-wave (mmWave) communication systems.


To the best of our knowledge, the integration of FAS and RSM has not been explored in the literature. Therefore, this paper proposes the FA-empowered RSM (FA-RSM) system to enhance the performance of RSM and bridge this research gap. In the proposed system, the transmitter deploys an FA and activates multiple ports simultaneously based on the designed port selection algorithms.
This study differs from previous antenna selection studies for traditional RSM systems in several key aspects. 
First, in the FA, a large number of ports are typically distributed within a confined area. The spatial correlation among these ports, along with the limited hardware deployment range, leads to a gradual saturation of the performance gain from activating additional ports, as confirmed by the simulation results.
Second, the port selection algorithms proposed in this paper are applicable under both ZF precoding and MMSE precoding. 
Furthermore, the inherent spatial correlation characteristics of FA are leveraged to develop a low-complexity port selection algorithm specifically tailored for the FA-RSM system. 
Additionally, theoretical tools for evaluating the performance of port selection are provided.
The main contributions of this paper are summarized as follows:
\begin{itemize}
    \item An innovative FA-empowered RSM system is developed. In this system, the transmitter is equipped with an FA that can simultaneously activate multiple ports, and the information is conveyed through modulation symbols and the indices of the receive antennas. The signal models for the FA-RSM system under both ZF and MMSE precoding are presented, along with the corresponding maximum likelihood detectors (MLD). 
    \item The optimal port selection algorithm is introduced based on capacity maximization. Subsequently, a low-complexity trace minimization-based decremental (TMD) port selection algorithm is proposed, leveraging the approximation of capacity under high SNR conditions and a greedy approach. To further reduce complexity, the maximum correlation elimination-assisted TMD (MCE-TMD) port selection algorithm is proposed, which exploits the spatial correlation characteristics of the FA. The complexities of the proposed algorithms are then thoroughly evaluated.
    \item The impact of port selection in FA-RSM systems is theoretically analyzed, and it is proven that the performance loss decreases as the number of activated ports increases. More explicitly, the capacity loss of under ZF precoding is derived, along with its upper bound. Additionally, the mean-square-error (MSE) expression under MMSE precoding is presented.
    \item A two-stage maximum energy-based detector (MED) is proposed to reduce the detection complexity. However, simulations reveal that the MED causes significant performance degradation in the FA-RSM system with MMSE precoding. To address this issue, a ratio threshold test-assisted detector (RTTD) is introduced as an effective trade-off between complexity and performance for the FA-RSM system with MMSE precoding.
    \item Simulation results show that: 1) With the assistance of FA, the proposed FA-RSM system significantly outperforms the traditional RSM system; 2) The performance gaps between the low-complexity port selection algorithms and the optimal port selection algorithm are relatively small; 3) Activating more ports improves the performance of the FA-RSM system, but with apparent diminishing returns; 4) The proposed low-complexity detectors achieve performance close to that of MLD detection.
\end{itemize}

The rest of this paper is organized as follows: Section~\ref{Sec-SystemModel} presents the FA-RSM system model. Section~\ref{Sec-PortSelect} provides the optimal port selection algorithm based on capacity maximization, along with the proposed low-complexity TMD and MCE-TMD algorithms. Section~\ref{Sec-PerformanceAnalysis} offers a theoretical analysis of the impact of port selection on the performance of the FA-RSM system. Section~\ref{Sec-DetectorDesign} elaborates on the proposed low-complexity MED and RTTD for detection. Section~\ref{Sec-Simulation} shows the simulation results. Section~\ref{Sec-Conclusion} concludes this paper.

\textit{Notations:} Scalar variables are denoted by italic letters, vectors are denoted by boldface small letters and matrices are denoted by boldface capital letters. $(\cdot)^*$, $(\cdot)^T$, $(\cdot)^{-1}$ and $(\cdot)^H$ denote the conjugate, transposition, inverse and Hermitian transposition operations, respectively. $\mathrm{tr}(\cdot)$ and $\mathrm{det}(\cdot)$ represent the trace and determinant of a matrix, respectively. $| \cdot |$ and ${\| \cdot \|}$ denote the absolute and the $\ell_2$ norm operations, respectively. $\binom{\cdot}{\cdot}$ denotes the binomial coefficient. $\mathrm{diag}(\cdot)$ denotes a diagonal matrix whose diagonal entries are the inputs. $\mathbf{x}(i)$, $\mathbf{X}_{[i,:]}$ and $\mathbf{X}_{[i,j]}$ denote the $i$-th entry of the vector $\mathbf{x}$, the $i$-th row of the matrix $\mathbf{X}$ and the entry in the $i$-th row and $j$-th column of the matrix $\mathbf{X}$, respectively. $\mathbb{E}[\cdot]$ returns the expectation of the input random quantity. $\langle \mathbf{x},\mathbf{y} \rangle$ returns the inner product of $\mathbf{x}$ and $\mathbf{y}$. The real and imaginary parts of a complex variable $X$ are denoted by $\Re\{X\}$ and $\Im\{X\}$.

\section{System Model} \label{Sec-SystemModel}
\begin{figure*}[t]
    \centerline{\includegraphics[width=0.9\textwidth]{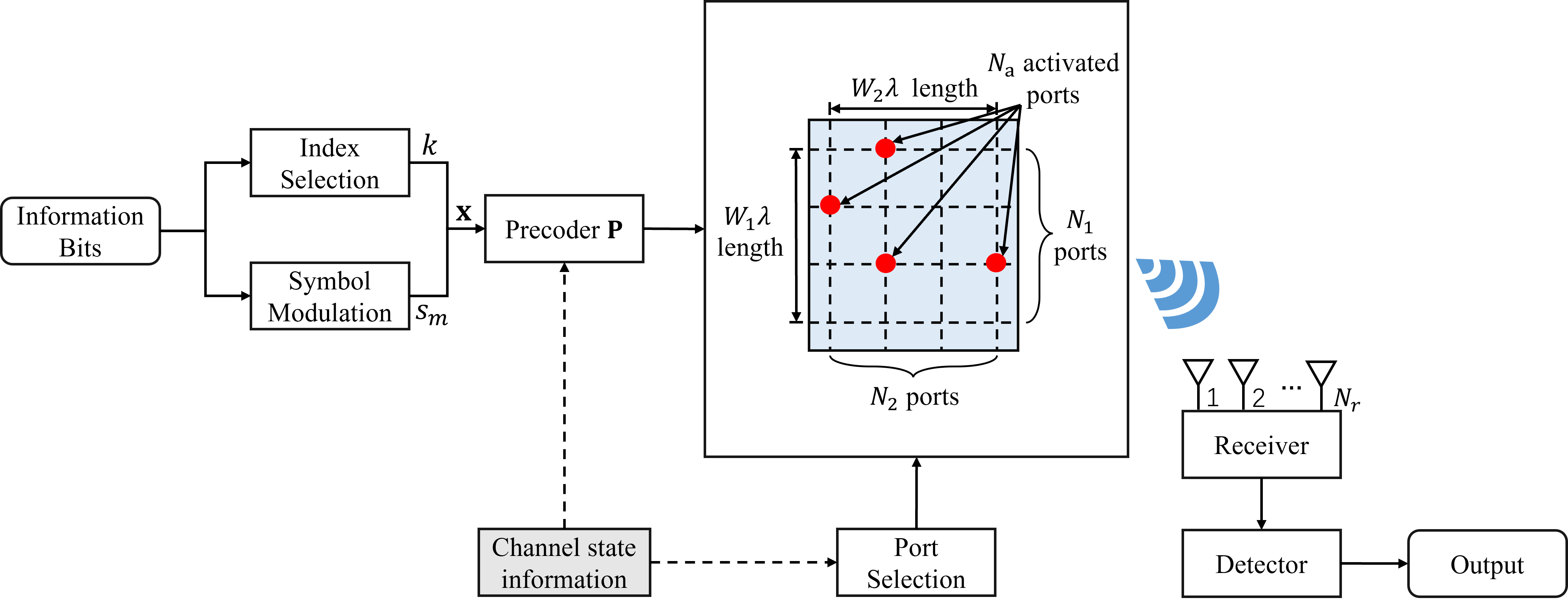}}
    
    \captionsetup{justification=raggedright, singlelinecheck=false}
    
    \caption{Block diagram of the proposed FA-RSM system.}
    \label{fig-SystemModel}
\end{figure*}
As depicted in \figref{fig-SystemModel}, consider a MIMO system equipped with $N_r$ conventional receive antennas, while the transmitter is equipped with an FA having $N$ ports. Assume that the FA occupies a two-dimensional (2D) space with an area of $W$. Consider a grid structure where $N_1$ and $N_2$ ports are uniformly distributed along the vertical and horizontal directions in linear spaces of lengths $\lambda W_1$ and $\lambda W_2$, respectively, so that $W=\lambda W_1 \times \lambda W_2$ and $N=N_1 \times N_2$, where $\lambda$ represents the wavelength of the carrier. In the FA-RSM system, the transmitter selects and activates $N_a$ from $N$ ports for signal transmission, where $N_a \geq N_r$ is assumed to ensure the existence of the right inverse of the channel matrix during precoding.

\subsection{Spatial Correlation Model}  \label{SubSec-SystemModel-CorrelationModel}
A 2D Cartesian coordinate system with the $x-O-y$ axes is established on the plane where the FA is located. 
Then, according to the channel model of the FAS in \cite{ref-MIMO-FAS}, the spatial correlation between the $n$-th port and the $\bar{n}$-th port, where $n, \bar{n} \in \{1,2,\ldots, N\}$, is given by 
\begin{equation}
\begin{aligned}
\label{eq-PortCorrelation}
J_{n,\bar{n}}&= j_0 \left(\frac{2 \pi}{\lambda} \sqrt{(x_n-x_{\bar{n}})^2+(y_n-y_{\bar{n}})^2} \right), 
\end{aligned}
\end{equation}%
where $j_0 (\cdot)$ is the spherical Bessel function of the first kind, and $(x_n, y_n)$ and $(x_{\bar{n}}, y_{\bar{n}})$ represent the coordinates of the $n$-th port and the $\bar{n}$-th port, respectively. 
The spatial correlation matrix at the transmitter can be represented as 
\begin{equation}
\label{eq-CorrelationMatrix}
\mathbf{J}_t=\left[
\begin{array}{cccc}
J_{1,1}&J_{1,2}&\ldots&J_{1,N}\\
J_{2,1}&J_{2,2}&\ldots&J_{2,N}\\
\vdots&\vdots&\ddots&\vdots\\
J_{N,1}&J_{N,2}&\ldots&J_{N,N}
\end{array}
\right].
\end{equation}%
As shown in \eqref{eq-PortCorrelation}, $J_{n,\bar{n}}= J_{\bar{n},n}$ always holds, so $\mathbf{J}_t$ is a real symmetric matrix that can be eigen-decomposed as $\mathbf{J}_t = \mathbf{U}_t \mathbf{\Lambda}_t \mathbf{U}_t^H$, where $\mathbf{U}_t$ is the matrix composed of the $N$ eigenvectors of $\mathbf{J}_t$, and $\mathbf{\Lambda}_t = \mathrm{diag}(\lambda_1^t, \ldots, \lambda_N^t)$ is a diagonal matrix with the corresponding eigenvalues of $\mathbf{J}_t$ on its diagonal. 

Due to the deployment of a conventional antenna array at the receiver and the focus of this paper on investigating the role and effect of the FA deployed at the transmitter in the RSM system, the spatial correlation between the receiver antennas is not considered in this paper. 
Therefore, the channel $\mathbf{H}=[\mathbf{h}_1,\mathbf{h}_2, \ldots, \mathbf{h}_N] \in \mathbb{C}^{N_r \times N}$ of the FA-RSM system can be modeled as 
\begin{equation}
\begin{aligned}
\label{eq-ChannelModel}
\mathbf{H}= \check{\mathbf{H}} \mathbf{J}_t^{\frac{1}{2}} = \check{\mathbf{H}} \sqrt{\mathbf{\Lambda}_t^H} \mathbf{U}_t^H,
\end{aligned}
\end{equation}%
where $\check{\mathbf{H}} \in \mathbb{C}^{N_r \times N}$ is a circularly symmetric complex Gaussian (CSCG) random matrix with each entry being independent and identically distributed (i.i.d.) and following $\mathcal{CN} (0,1)$. 

\subsection{Signal Model}  \label{SubSec-SystemModel-SignalModel}
Based on the concept of RSM, within each transmission slot, the input information bits are split into two parts. The first part is mapped to a modulation symbol, denoted by $s_m$ for $m \in \{1, 2, \ldots, M\}$, which belongs to a normalized $M$-ary constellation, with $\mathbb{E}[|s_m|^2]=1$. The second part is mapped to an index of the receive antenna, denoted by $k$ for $k \in \{1, 2, \ldots, N_r\}$. More specifically, the information bits are mapped to the spatial modulation vector 
\begin{equation}
\label{eq-SM-vector}
\mathbf{x}= s_m \mathbf{e}_k,
\end{equation}%
where $\mathbf{e}_k$ represents the $k$-th column of the identity matrix $\mathbf{I}_{N_r}$ of size $N_r$. Therefore, the SE of FA-RSM, in terms of bits per channel use (bpcu), is given by
\begin{equation}
\label{eq-SE-FA-RSM}
\mathrm{SE_{FA-RSM}}= \log_2 M + \log_2 N_r \ [\mathrm{bpcu}].
\end{equation}%

Subsequently, the modulation vector $\mathbf{x}$ is multiplied by the precoding matrix $\mathbf{P}=[\mathbf{p}_1,\mathbf{p}_2, \ldots, \mathbf{p}_{N_r}] \in \mathbb{C}^{N_a \times N_r}$ for transmission. To maintain a constant average transmit power, i.e., $\mathbb{E}[\|\mathbf{P} \mathbf{x}\|^2]=1$, the precoding matrix $\mathbf{P}$ of FA-RSM must satisfy the constraint $\mathrm{tr}(\mathbf{P}\mathbf{P}^H)= N_r$. 

Denote the effective channel matrix after port selection as $\tilde{\mathbf{H}}= [\tilde{\mathbf{h}}_1,\tilde{\mathbf{h}}_2, \ldots, \tilde{\mathbf{h}}_{N_a}] \in \mathbb{C}^{N_r \times N_a}$, which is formed by the $N_a$ columns of $\mathbf{H}$ (the algorithms for port selection are described in detail in Section \ref{Sec-PortSelect}). Ignoring path loss, the received signal $\mathbf{y}=[y_1, y_2, \ldots, y_{N_r}]^T$ can be expressed as 
\begin{equation}
\label{eq-ReceivedSignal}
\mathbf{y}= \tilde{\mathbf{H}} \mathbf{P} \mathbf{x} + \mathbf{w},
\end{equation}%
where 
$\mathbf{w} \in \mathbb{C}^{N_r \times 1}$ is the additive white Gaussian noise (AWGN) vector with each entry $w_k$ being i.i.d. and following $\mathcal{CN} (0,N_0)$. 

\subsection{Precoding}  \label{SubSec-SystemModel-Precoding}
The precoding matrix should maximize the energy at the intended receive antenna as much as possible, while simultaneously minimizing energy leakage to other unintended receive antennas. To maintain system simplicity, this paper considers two classical linear precoding schemes: ZF precoding and MMSE precoding.

For ZF precoding, we have 
\begin{equation}
\label{eq-ZFprecoding}
\mathbf{P}= \beta_{\mathrm{ZF}} \tilde{\mathbf{H}}^H (\tilde{\mathbf{H}} \tilde{\mathbf{H}}^H)^{-1},
\end{equation}%
where $\beta_{\mathrm{ZF}} = \sqrt{N_r / \mathrm{tr} \left( (\tilde{\mathbf{H}} \tilde{\mathbf{H}}^H)^{-1} \right) }$ is the normalized factor. Assuming that the spatial modulation vector $\mathbf{x}= s_m \mathbf{e}_k$ is transmitted, substituting \eqref{eq-ZFprecoding} into \eqref{eq-ReceivedSignal} yields the expression of the received signal as 
\begin{equation}
\begin{aligned}
\label{eq-ReceivedSignal-ZF}
    y_k&= \beta_{\mathrm{ZF}} s_m + w_k,\\
    y_{\bar{k}}&= w_{\bar{k}}, \ {\bar{k}} \in \{1,2, \ldots, N_r\} \backslash \{k\}.
\end{aligned}
\end{equation}%
It can be observed from \eqref{eq-ReceivedSignal-ZF} that, with the assistance of ZF precoding, information can only be obtained at the $k$-th receive antenna, while all other unintended receive antennas only contain noise.

By contrast, for MMSE precoding, we have
\begin{equation}
\label{eq-MMSEprecoding}
\mathbf{P}= \beta_{\mathrm{MMSE}} \tilde{\mathbf{H}}^H (\tilde{\mathbf{H}} \tilde{\mathbf{H}}^H + N_r N_0 \mathbf{I}_{N_r})^{-1},
\end{equation}%
where $\beta_{\mathrm{MMSE}} = \sqrt{N_r / \mathrm{tr} \left( \tilde{\mathbf{H}} \tilde{\mathbf{H}}^H (\tilde{\mathbf{H}} \tilde{\mathbf{H}}^H + N_r N_0 \mathbf{I}_{N_r})^{-2} \right) }$. 
Similarly, assuming $\mathbf{x}= s_m \mathbf{e}_k$ is transmitted, the received signal can be further written as \begin{equation}
\begin{aligned}
\label{eq-ReceivedSignal-MMSE}
    y_k= \beta_{\mathrm{MMSE}} \sum_{l=1}^{N_r} \tilde{\mathbf{H}}_{[k,:]} \tilde{\mathbf{H}}_{[l,:]}^H (\tilde{\mathbf{H}} \tilde{\mathbf{H}}^H + N_r N_0 \mathbf{I}_{N_r})^{-1}_{[l,k]} s_m &+ w_k,\\
    y_{\bar{k}}= \beta_{\mathrm{MMSE}} \sum_{l=1}^{N_r} \tilde{\mathbf{H}}_{[\bar{k},:]} \tilde{\mathbf{H}}_{[l,:]}^H (\tilde{\mathbf{H}} \tilde{\mathbf{H}}^H + N_r N_0 \mathbf{I}_{N_r})^{-1}_{[l,k]} s_m &+ w_{\bar{k}}, \\
    {\bar{k}} \in \{1,2, \ldots, N_r\} &\backslash \{k\}.
\end{aligned}
\end{equation}%
Unlike ZF precoding, MMSE precoding fully accounts for the impact of noise on the received signal, thereby preventing the power attenuation of effective signal components under certain channel conditions.

\subsection{Maximum Likelihood Detection}  \label{SubSec-SystemModel-Detection}
The optimal MLD of FA-RSM can be expressed as 
\begin{equation}
\begin{aligned}
\label{eq-MLD}
    (\hat{m},\hat{k})&= \arg \underset{m,k}{\min} \left \| \mathbf{y} - \tilde{\mathbf{H}} \mathbf{P} \mathbf{x} \right \|^2\\
    &=\arg \underset{m,k}{\min} \left \| \mathbf{y} - \tilde{\mathbf{H}} \mathbf{P} \mathbf{e}_k s_m \right \|^2.
\end{aligned}
\end{equation}%

When ZF precoding is employed, substituting \eqref{eq-ZFprecoding} into \eqref{eq-MLD} simplifies the MLD to
\begin{equation}
\label{eq-MLD-ZF}
    (\hat{m},\hat{k}) = \arg \underset{m,k}{\min} \left \| \mathbf{y} - \beta_{\mathrm{ZF}} s_m \mathbf{e}_k \right \|^2.
\end{equation}%
When MMSE precoding is employed, substituting \eqref{eq-MMSEprecoding} into \eqref{eq-MLD} rewrites the MLD as
\begin{equation}
\begin{aligned}
\label{eq-MLD-MMSE}
    (\hat{m},\hat{k}) &= \arg \underset{m,k}{\min} \left \| \mathbf{y} - \beta_{\mathrm{MMSE}} \mathbf{G} \mathbf{e}_k s_m \right \|^2 \\
    &= \arg \underset{m,k}{\min} \left \| \mathbf{y} - \beta_{\mathrm{MMSE}} s_m \mathbf{g}_k \right \|^2 ,
\end{aligned}
\end{equation}%
where $\mathbf{G} = [\mathbf{g}_1,\mathbf{g}_2, \ldots, \mathbf{g}_{N_r}] = \tilde{\mathbf{H}} \tilde{\mathbf{H}}^H (\tilde{\mathbf{H}} \tilde{\mathbf{H}}^H + N_r N_0 \mathbf{I}_{N_r})^{-1}$. Eqs.~\eqref{eq-MLD-ZF} and \eqref{eq-MLD-MMSE} together reveal that the RSM system, with the assistance of precoding, reduces the computational complexity of detection, making it a receiver-friendly transmission scheme. 
However, the optimal MLD exhaustively searches the possible spatial modulation vectors $\mathbf{x}$, which is highly time-consuming, with a time complexity of $\mathcal{O}(M N_r)$. Therefore, this paper further proposes two low-complexity detectors, which are described in detail in Section \ref{Sec-DetectorDesign}.

\section{Port Selection Algorithms}  \label{Sec-PortSelect}
Based on the proposed FA-RSM system model, port selection plays a crucial role. Therefore, this section first introduces the optimal port selection algorithm based on capacity maximization. Leveraging ZF and MMSE precoding matrices employed in the FA-RSM system, we formulate an approximate criterion derived from the capacity-based metric. Building on this, the TMD algorithm is developed following the greedy principle. To further reduce complexity, the MCE-TMD algorithm is proposed utilizing the spatial correlation characteristics of the FA. In contrast to existing MIMO-FAS systems, the proposed low-complexity port selection algorithms are custom-designed based on the simple precoding techniques employed in the FA-RSM system, making them both simpler and more targeted. 
The simulation results presented in Section \ref{Sec-Simulation} further confirm that, thanks to the transmission mechanism of the FA-RSM system, even simple designs can achieve performance comparable to that of the optimal algorithm.
Finally, a detailed complexity analysis of the aforementioned port selection algorithms is provided.

\subsection{Optimal Capacity Maximization-Based Port Selection} \label{Sec-PortSelect-Optimal}

Let $\mathcal{I}$ denote the set consisting of the indices of $N_a$ ports, i.e., $|\mathcal{I}| = N_a$ and $\mathcal{I} \subset \{1, 2, \ldots, N\}$. Therefore, the set $\mathcal{I}$ has $S = \binom{N}{N_a}$ possible selections. 
Let $\mathbf{H}_{\mathcal{I}}$ denote the submatrix of $\mathbf{H}$, obtained by selecting $N_a$ columns from $\mathbf{H}$ according to the $N_a$ indices in the set $\mathcal{I}$.
Based on \cite{ref-MIMO-FAS}, given $\mathbf{H}_{\mathcal{I}}$, the capacity of the FA-RSM system can be calculated by
\begin{equation}
\label{eq-Capacity}
    C_{\mathcal{I}} = \log_2 \mathrm{det} \left( \mathbf{I}_{N_r} + \frac{1}{N_r N_0} \mathbf{H}_{\mathcal{I}} \mathbf{P}_{\mathcal{I}} \mathbf{P}_{\mathcal{I}}^H \mathbf{H}_{\mathcal{I}}^H \right),
\end{equation}%
where $\mathbf{P}_{\mathcal{I}}$ is the precoding matrix corresponding to the matrix $\mathbf{H}_{\mathcal{I}}$, and the coefficient $\frac{1}{N_r}$, unlike in MIMO-FAS, arises from the power normalization constraint of the precoding in the FA-RSM system.
Therefore, the optimal port set $\mathcal{I}^*$ can be obtained based on the capacity maximization criterion from 
\begin{equation}
\begin{aligned}
\label{eq-CapacityMaximize}
    \mathcal{I}^* &= \arg \underset{\mathcal{I}}{\max} \ C_{\mathcal{I}} \\
    &= \arg \underset{\mathcal{I}}{\max} \ \sum_{k=1}^{N_r} \log_2 \left( 1 + \frac{(\sigma_k^{\mathcal{I}})^2}{N_r N_0} \right),
\end{aligned}
\end{equation}%
where $\sigma_k^{\mathcal{I}}$ represents the $k$-th singular value of the matrix $\mathbf{H}_{\mathcal{I}} \mathbf{P}_{\mathcal{I}}$. 
As can be seen in \eqref{eq-CapacityMaximize}, the optimal port selection algorithm requires exhaustive search of $S$ possible port selections, and for each selection, the singular values of the corresponding matrix $\mathbf{H}_{\mathcal{I}} \mathbf{P}_{\mathcal{I}}$ need to be computed, resulting in high computational complexity. To address this, we propose the following fast decremental port selection algorithm.

\subsection{TMD Port Selection} \label{Sec-PortSelect-TMD}

Consider the scenario with a high signal-to-noise ratio (SNR), i.e., $N_0 \to 0$. In this context, \eqref{eq-CapacityMaximize} can be approximated as 
\begin{equation}
\begin{aligned}
\label{eq-CapacityMaximizeApprox}
    \mathcal{I}^* &= \arg \underset{\mathcal{I}}{\max} \ \log_2 \mathrm{det} \left( \mathbf{I}_{N_r} + \frac{1}{N_r N_0} \mathbf{H}_{\mathcal{I}} \mathbf{P}_{\mathcal{I}} \mathbf{P}_{\mathcal{I}}^H \mathbf{H}_{\mathcal{I}}^H \right) \\
    &\approx \arg \underset{\mathcal{I}}{\max} \ \mathrm{det} \left(\mathbf{H}_{\mathcal{I}} \mathbf{P}_{\mathcal{I}} \mathbf{P}_{\mathcal{I}}^H \mathbf{H}_{\mathcal{I}}^H \right).
\end{aligned}
\end{equation}%

At this point, for ZF precoding, $\mathbf{H}_{\mathcal{I}} \mathbf{P}_{\mathcal{I}} = \beta_{\mathrm{ZF}} \mathbf{H}_{\mathcal{I}} \mathbf{H}_{\mathcal{I}}^H (\mathbf{H}_{\mathcal{I}} \mathbf{H}_{\mathcal{I}}^H)^{-1} = \sqrt{N_r / \mathrm{tr} \left( (\mathbf{H}_{\mathcal{I}} \mathbf{H}_{\mathcal{I}}^H)^{-1} \right) }\mathbf{I}_{N_r}$. 
For MMSE precoding, $\mathbf{H}_{\mathcal{I}} \mathbf{P}_{\mathcal{I}} = \beta_{\mathrm{MMSE}} \mathbf{H}_{\mathcal{I}} \mathbf{H}_{\mathcal{I}}^H (\mathbf{H}_{\mathcal{I}} \mathbf{H}_{\mathcal{I}}^H + N_r N_0 \mathbf{I}_{N_r})^{-1} \approx \sqrt{N_r / \mathrm{tr} \left( (\mathbf{H}_{\mathcal{I}} \mathbf{H}_{\mathcal{I}}^H)^{-1} \right) }\mathbf{I}_{N_r}$, since $N_0 \to 0$ in this scenario. 
Therefore, whether ZF precoding or MMSE precoding is employed, in the high-SNR scenario, \eqref{eq-CapacityMaximizeApprox} can be approximated as
\begin{equation}
\begin{aligned}
\label{eq-TraceMinimize}
    \mathcal{I}^* &= \arg \underset{\mathcal{I}}{\max} \ \mathrm{det} \left( \frac{N_r}{\mathrm{tr} \left( (\mathbf{H}_{\mathcal{I}} \mathbf{H}_{\mathcal{I}}^H)^{-1} \right)} \mathbf{I}_{N_r} \right) \\
    &= \arg \underset{\mathcal{I}}{\min} \ \mathrm{tr} \left( (\mathbf{H}_{\mathcal{I}} \mathbf{H}_{\mathcal{I}}^H)^{-1} \right) .
\end{aligned}
\end{equation}%
The problem has been transformed into finding a port index set $\mathcal{I}$ that minimizes the trace of the matrix $(\mathbf{H}_{\mathcal{I}} \mathbf{H}_{\mathcal{I}}^H)^{-1}$. 
However, the computational complexity of \eqref{eq-TraceMinimize} remains relatively high, as an exhaustive search over $S$ possible port selections is still required, with the inverse of the corresponding matrix $\mathbf{H}_{\mathcal{I}} \mathbf{H}_{\mathcal{I}}^H$ computed for each selection. 

Hence, iteration is introduced in the proposed algorithm to substitute the exhaustive search. Before reaching the final iterative algorithm, the necessary symbol notations and prerequisites are introduced as follows.
Let $\mathbf{H}_{-i} = [\mathbf{h}_1, \dots, \mathbf{h}_{i-1}, \mathbf{h}_{i+1}, \ldots, \mathbf{h}_N] \in \mathbb{C}^{N_r \times (N-1)}$ denote the submatrix of $\mathbf{H}$ obtained by removing the $i$-th column $\mathbf{h}_i$ from $\mathbf{H}$. 
According to block matrix multiplication, $\mathbf{H}_{-i} \mathbf{H}_{-i}^H = \mathbf{H} \mathbf{H}^H - \mathbf{h}_i \mathbf{h}_i^H$ holds. Moreover, according to the Sherman-Morrison-Woodbury (SMW) formula \cite{ref-SMW}, the following equality holds:
\begin{equation}
\label{eq-SMW}
    (\mathbf{Z} + \mathbf{X}\mathbf{Y})^{-1}= \mathbf{Z}^{-1} - \mathbf{Z}^{-1} \mathbf{X} (\mathbf{I} + \mathbf{Y} \mathbf{Z}^{-1} \mathbf{X})^{-1} \mathbf{Y} \mathbf{Z}^{-1}.
\end{equation}%
Utilizing the SMW formula, we can obtain
\begin{equation}
\begin{aligned}
\label{eq-SMW-H}
    (\mathbf{H}_{-i} \mathbf{H}_{-i}^H)^{-1} &= (\mathbf{H} \mathbf{H}^H - \mathbf{h}_i \mathbf{h}_i^H)^{-1} \\
    &= (\mathbf{H} \mathbf{H}^H)^{-1} + \frac{(\mathbf{H} \mathbf{H}^H)^{-1} \mathbf{h}_i \mathbf{h}_i^H (\mathbf{H} \mathbf{H}^H)^{-1}}{1 - \mathbf{h}_i^H (\mathbf{H} \mathbf{H}^H)^{-1} \mathbf{h}_i}.
\end{aligned}
\end{equation}%

Now, consider an iterative decremental method to remove $N - N_a$ ports from the $N$ ports. Let $i^*_t$ denote the index of the port removed in the $t$-th iteration. For the first iteration, one port needs to be removed from the $N$ ports. In this context, based on the concept of the greedy algorithm, the port selection problem can be transformed from \eqref{eq-TraceMinimize} to 
\begin{equation}
\label{eq-TraceMinimize-i}
    i^*_1 = \arg \underset{i \in \mathcal{I}_0}{\min} \ \mathrm{tr} \left( (\mathbf{H}_{-i} \mathbf{H}_{-i}^H)^{-1} \right),
\end{equation}%
where $\mathcal{I}_0=\{1, 2, \ldots, N\}$ denotes the set of indices of all $N$ ports. Then, with the help of \eqref{eq-SMW-H}, \eqref{eq-TraceMinimize-i} can be further written as
\begin{equation}
\begin{aligned}
\label{eq-TraceMinimize-One}
    i^*_1 &= \arg \underset{i \in \mathcal{I}_0}{\min} \ \mathrm{tr} \left( \mathbf{A} + \frac{\mathbf{A} \mathbf{h}_i \mathbf{h}_i^H \mathbf{A}}{1 - \mathbf{h}_i^H \mathbf{A} \mathbf{h}_i} \right) \\
    &= \arg \underset{i \in \mathcal{I}_0}{\min} \ \mathrm{tr} \left( \frac{\mathbf{A} \mathbf{h}_i \mathbf{h}_i^H \mathbf{A}}{1 - \mathbf{h}_i^H \mathbf{A} \mathbf{h}_i} \right) \\
    &= \arg \underset{i \in \mathcal{I}_0}{\min} \ \frac{\| \mathbf{A} \mathbf{h}_i \|}{1 - \mathbf{h}_i^H \mathbf{A} \mathbf{h}_i},
\end{aligned}
\end{equation}%
where $\mathbf{A} = (\mathbf{H} \mathbf{H}^H)^{-1}$. The second equality of \eqref{eq-TraceMinimize-One} utilizes the property of the matrix trace, $\mathrm{tr}(\mathbf{A} + \mathbf{X}) = \mathrm{tr}(\mathbf{A}) + \mathrm{tr}(\mathbf{X})$, along with the fact that $\mathrm{tr}(\mathbf{A})$ remains unchanged for any $i \in \mathcal{I}_0$. The final equality of \eqref{eq-TraceMinimize-One} utilizes the fact that $\mathbf{A} = \mathbf{A}^H$.
It is worth noting that, compared to the process of solving \eqref{eq-TraceMinimize-i}, which requires N inverse operations, the process of solving \eqref{eq-TraceMinimize-One} only requires a single inverse operation, i.e., computing $\mathbf{A}$. 

Then, for the second iteration, one port needs to be excluded from the remaining $N - 1$ ports. With the selection result $i^*_1$ from the first iteration, using the same derivation process, we can obtain
\begin{equation}
\label{eq-TraceMinimize-Two}
    i^*_2 = \arg \underset{i \in \mathcal{I}_0 \backslash \{i^*_1\}}{\min} \ \frac{\| \mathbf{A}^\prime \mathbf{h}_i \|}{1 - \mathbf{h}_i^H \mathbf{A}^\prime \mathbf{h}_i},
\end{equation}%
in which $\mathbf{A}^\prime = (\mathbf{H}_{-i^*_1} \mathbf{H}_{-i^*_1}^H)^{-1}$. 
Notably, matrix $\mathbf{A}^\prime$ can now be computed without matrix inversion by directly substituting $i^*_1$ and $\mathbf{A}$, obtained from the first iteration, into \eqref{eq-SMW-H} as
\begin{equation}
\label{eq-UpdateB}
\mathbf{A}^\prime = \mathbf{A} + \frac{\mathbf{A} \mathbf{h}_{i^*_1} \mathbf{h}_{i^*_1}^H \mathbf{A}}{1 - \mathbf{h}_{i^*_1}^H \mathbf{A} \mathbf{h}_{i^*_1}}.
\end{equation}%

Each subsequent iteration can follow the above procedure to remove ports and simultaneously update the matrix, until the iteration terminates with $N_a$ ports remaining. The proposed iterative port selection algorithm, referred to as the TMD algorithm, is summarized in \textbf{Algorithm \ref{alg-TMD}}.
\begin{algorithm}[t] 
\setcounter{algorithm}{0}   
\caption{Proposed TMD port selection algorithm}
\label{alg-TMD} 
\renewcommand{\algorithmicrequire}{\textbf{Input:}}
\renewcommand{\algorithmicensure}{\textbf{Output:}}
\begin{algorithmic}[1]
    \REQUIRE $\mathbf{H}$, $N$, $N_a$.
    \ENSURE $\tilde{\mathbf{H}}$, $\mathcal{I}^*$. 
    \STATE Initialization: $\mathcal{I} = \{1, 2, \ldots, N\}$, $\mathbf{A} = (\mathbf{H} \mathbf{H}^H)^{-1}$.
    \FOR{$t=1:N-N_a$} 
        \STATE $i^* \gets \arg \underset{i \in \mathcal{I}}{\min} \ \frac{\| \mathbf{A} \mathbf{h}_i \|}{1 - \mathbf{h}_i^H \mathbf{A} \mathbf{h}_i}$;
        \STATE $\mathbf{A} \gets \mathbf{A} + \frac{\mathbf{A} \mathbf{h}_{i^*} \mathbf{h}_{i^*}^H \mathbf{A}}{1 - \mathbf{h}_{i^*}^H \mathbf{A} \mathbf{h}_{i^*}} $;
        \STATE $\mathcal{I} \gets \mathcal{I} \backslash \{i^*\}$;
    \ENDFOR
    \STATE $\mathcal{I}^* \gets \mathcal{I}$;
    \STATE $\tilde{\mathbf{H}} \gets \mathbf{H}_{\mathcal{I}^*}$;
\end{algorithmic} 
\end{algorithm}

\subsection{MCE-TMD Port Selection} \label{Sec-PortSelect-MCE-TMD}

For the FAS, the port number $N$ is typically substantial, leading to a relatively high computational complexity of the proposed TMD algorithm, as examined in the subsequent Section \ref{Sec-PortSelect-Complexity}. 
Therefore, based on the TMD algorithm, we further propose a port selection algorithm to reduce the complexity. This algorithm utilizes the high spatial correlation among densely distributed ports to perform a pre-selection, referred to as the MCE-TMD algorithm.

First, the spatial correlation values among the $N$ ports are sorted. According to \eqref{eq-PortCorrelation} and $J_{n,\bar{n}}= J_{\bar{n},n}$, there are a total of $S^{\prime} = \binom{N}{2} = \frac{N(N-1)}{2}$ values arranged in descending order as: $J_{n_1,\bar{n}_1} \geq \ldots \geq J_{n_j,\bar{n}_j} \geq \ldots \geq J_{ n_{S^{\prime}},\bar{n}_{S^{\prime}} }$, where $1 \leq n_j < \bar{n}_j \leq N$, and $J_{n_j,\bar{n}_j}$ represents the $j$-th largest spatial correlation value, which is generated between the $n_j$-th and $\bar{n}_j$-th ports.
Then, the sorted index values are stored in the arrays 
\begin{equation}
\begin{aligned}
\label{eq-SortCorrIndex}
    \mathbf{n}_0&= [n_1, n_2, \ldots, n_{S^{\prime}}], \\
    \bar{\mathbf{n}}_0&= [\bar{n}_1, \bar{n}_2, \ldots, \bar{n}_{S^{\prime}}],
\end{aligned}
\end{equation}%
respectively. 
Notably, the order of spatial correlation among ports only depends on the pre-set port distribution conditions. Once the configuration of the FA-RSM system is given, \eqref{eq-SortCorrIndex} remains constant. Even if the channel $\mathbf{H}$ changes and the port selection algorithm needs to be re-executed, there is no need to recompute \eqref{eq-SortCorrIndex}. 

Let $N_b$ ($N > N_b > N_a$) denote the number of remaining ports after pre-selection. 
As described in \textbf{Algorithm \ref{alg-MCE-TMD}}, the proposed MCE-TMD algorithm consists of two main stages. 
The first stage iteratively excludes $N - N_b$ ports from the $N$ ports based on MCE. 
Specifically, during each iteration, the top $N_b$ (for simplicity) pairwise combinations of the remaining ports with higher correlation are selected. 
The inner product of the two columns $\mathbf{h}_{\mathbf{n}(b)}$ and $\mathbf{h}_{\bar{\mathbf{n}}(b)}$ corresponding to each port pair $\mathbf{n}(b)$ and $\bar{\mathbf{n}}(b)$, where $b=1,2,\ldots,N_b$, is then calculated.
The two columns $\mathbf{h}_{\mathbf{n}(b^*)}$ and $\mathbf{h}_{\bar{\mathbf{n}}(b^*)}$ that yield the maximum inner product are subsequently identified. 
The port index $i^*$ to be excluded in the current iteration is the one corresponding to the column with the smaller Euclidean norm between $\mathbf{h}_{\mathbf{n}(b^*)}$ and $\mathbf{h}_{\bar{\mathbf{n}}(b^*)}$.
Update the arrays $\mathbf{n}$ and $\bar{\mathbf{n}}$, which store the correlation ranking indices, by removing the port pair that contains the $i^*$-th port.
Next, with the port pre-selection results from the first stage, the second stage executes the TMD algorithm to select $N_a$ ports from the remaining $N_b$ ports.

\begin{algorithm}[t] 
\setcounter{algorithm}{1}   
\caption{Proposed MCE-TMD port selection algorithm}
\label{alg-MCE-TMD} 
\renewcommand{\algorithmicrequire}{\textbf{Input:}}
\renewcommand{\algorithmicensure}{\textbf{Output:}}
\begin{algorithmic}[1]
    \REQUIRE $\mathbf{H}$, $\mathbf{n}_0$, $\bar{\mathbf{n}}_0$, $N$, $N_b$, $N_a$.
    \ENSURE $\tilde{\mathbf{H}}$, $\mathcal{I}^*$. 
    \STATE Initialization: $\mathcal{I} = \{1, 2, \ldots, N\}$, $\mathbf{n}=\mathbf{n}_0$, $\bar{\mathbf{n}}=\bar{\mathbf{n}}_0$.
    \FOR{$t=1:N-N_b$} 
        \STATE $b^* \gets \arg \underset{b = 1,2,\ldots,N_b}{\max} \ \langle \mathbf{h}_{\mathbf{n}(b)},\mathbf{h}_{\bar{\mathbf{n}}(b)} \rangle$;
        \IF{$\| \mathbf{h}_{\mathbf{n}(b^*)} \| > \| \mathbf{h}_{\bar{\mathbf{n}}(b^*)} \|$}
            \STATE $i^* \gets \bar{\mathbf{n}}(b^*)$; 
            \STATE $\mathcal{J} \gets \{j \mid 1 \leq j \leq |\bar{\mathbf{n}}|, \bar{\mathbf{n}}(j) \neq i^*\}$;
        \ELSE
            \STATE $i^* \gets \mathbf{n}(b^*)$;
            \STATE $\mathcal{J} \gets \{j \mid 1 \leq j \leq |\mathbf{n}|, \mathbf{n}(j) \neq i^*\}$;
        \ENDIF
        \STATE $\mathbf{n} \gets \mathbf{n}_{\mathcal{J}}$;
        \STATE $\bar{\mathbf{n}} \gets \bar{\mathbf{n}}_{\mathcal{J}}$;
        \STATE $\mathcal{I} \gets \mathcal{I} \backslash \{i^*\}$;
    \ENDFOR
    \STATE $\mathbf{A} \gets (\mathbf{H}_\mathcal{I} \mathbf{H}_\mathcal{I}^H)^{-1}$;
    \FOR{$t=1:N_b-N_a$} 
        \STATE $i^* \gets \arg \underset{i \in \mathcal{I}}{\min} \ \frac{\| \mathbf{A} \mathbf{h}_i \|}{1 - \mathbf{h}_i^H \mathbf{A} \mathbf{h}_i}$;
        \STATE $\mathbf{A} \gets \mathbf{A} + \frac{\mathbf{A} \mathbf{h}_{i^*} \mathbf{h}_{i^*}^H \mathbf{A}}{1 - \mathbf{h}_{i^*}^H \mathbf{A} \mathbf{h}_{i^*}} $;
        \STATE $\mathcal{I} \gets \mathcal{I} \backslash \{i^*\}$;
    \ENDFOR
    \STATE $\mathcal{I}^* \gets \mathcal{I}$;
    \STATE $\tilde{\mathbf{H}} \gets \mathbf{H}_{\mathcal{I}^*}$;
\end{algorithmic} 
\end{algorithm}

\subsection{Complexity Analysis}  \label{Sec-PortSelect-Complexity}
For the optimal port selection algorithm in Section \ref{Sec-PortSelect-Optimal}, it is necessary to exhaustively search the $S = \binom{N}{N_a}$ port sets. 
The computational cost of each search mainly consists of two parts: first, computing the product of $\mathbf{H}_{\mathcal{I}}$ and $\mathbf{P}_{\mathcal{I}}$, and second, computing the singular values of $\mathbf{H}_{\mathcal{I}} \mathbf{P}_{\mathcal{I}}$ and further computing the capacity $C_{\mathcal{I}}$.
The complexity of computing $\mathbf{H}_{\mathcal{I}} \mathbf{P}_{\mathcal{I}}$ is $\mathcal{O}(N_a N_r^2)$, while the complexity of computing the singular values and the capacity is $\mathcal{O}(N_r^3)$. 
Therefore, the complexity order of the optimal algorithm is
\begin{equation}
\begin{aligned}
\label{eq-ComplexityOptimal}
\mathcal{O}(S(N_a N_r^2 + N_r^3)) &= \mathcal{O}\left( \frac{N!}{N_a! (N-N_a)!}(N_a N_r^2 + N_r^3) \right) \\
&\approx \mathcal{O}\left( \frac{N^{N_a}}{N_a!}(N_a N_r^2 + N_r^3) \right).
\end{aligned}
\end{equation}%

For the proposed TMD algorithm in Section \ref{Sec-PortSelect-TMD}, the matrix $\mathbf{A}$ needs to be calculated first, with a complexity order of $\mathcal{O}(N N_r^2 + N_r^3)$. 
Next, a total of $N - N_a$ iterations are performed. In the $t$-th iteration, the the metrics for $N - t + 1$ ports need to be calculated for decision-making, with a complexity order of $\mathcal{O}(N_r^2)$, and the complexity of updating matrix $\mathbf{A}$ is $\mathcal{O}(N_r^2)$. 
Therefore, the complexity order of the TMD algorithm is 
\begin{equation}
\begin{aligned}
\label{eq-ComplexityTMD}
&\mathcal{O}\left( N N_r^2 + N_r^3 + N_r^2\sum_{t=1}^{N-N_a} (N - t + 1) + (N - N_a)N_r^2 \right) \\
&\approx \mathcal{O}\left( (N^2 - N_a^2) N_r^2 \right).
\end{aligned}
\end{equation}%

For the proposed MCE-TMD algorithm in Section \ref{Sec-PortSelect-MCE-TMD}, the first stage pre-selects $N_b$ ports, requiring $N - N_b$ iterations. 
In each iteration, the complexity of calculating the $N_b$ inner products is $\mathcal{O}(N_b N_r)$, and the complexity of comparing the Euclidean norms between two columns is $\mathcal{O}(N_r)$. Therefore, the complexity of the first stage is $\mathcal{O}((N - N_b) N_b N_r)$. 
Utilizing the previously analyzed complexity \eqref{eq-ComplexityTMD} of the TMD algorithm, the complexity of the second stage is $\mathcal{O}((N_b^2 - N_a^2) N_r^2)$. Therefore, the complexity order of the MCE-TMD algorithm is 
\begin{equation}
\begin{aligned}
\label{eq-ComplexityMCETMD}
&\mathcal{O}\left((N - N_b) N_b N_r + (N_b^2 - N_a^2) N_r^2 \right) \\
&\approx \mathcal{O}\left((N - N_b) N_b N_r\right),
\end{aligned}
\end{equation}%
Compared to \eqref{eq-ComplexityTMD}, it can be seen that the MCE-TMD algorithm further reduces the complexity based on the TMD algorithm, highlighting the advantage of the MCE-TMD algorithm.

\section{Theoretical Analysis of Port Selection Impact}  \label{Sec-PerformanceAnalysis}
This section provides a theoretical analysis of port selection in the FA-RSM system, demonstrating that increasing the number of activated ports improves system performance. Specifically, we provide the capacity loss formula and its upper bound under ZF precoding, and present the MSE formula under MMSE precoding.

\subsection{Analysis of ZF Precoding} \label{Sec-PerformanceAnalysis-ZF}
Let $\hat{\mathcal{I}}$ denote a port index set that is larger than the index set of the activated ports $\mathcal{I}$, i.e., $\mathcal{I} \subset \hat{\mathcal{I}} \subseteq \mathcal{I}_0 = \{1, 2, \ldots, N\}$, and let $\bar{\mathcal{I}} = \hat{\mathcal{I}} \backslash \mathcal{I}$ represent the set difference between these two sets.
According to the Theorem in \cite{ref-TAS-PrecodedMIMO}, for sets $\mathcal{I}$ and $\hat{\mathcal{I}}$ with $|\mathcal{I}| < |\hat{\mathcal{I}}|$, it follows that $C_{\mathcal{I}} < C_{\hat{\mathcal{I}}}$. This indicates that the greater the number of activated ports, the higher the capacity of the FA-RSM system.

Then, by substituting the ZF precoding matrix \eqref{eq-ZFprecoding}, the capacity difference $C_d$ between the two sets, regarded as the capacity loss, can be expressed as
\begin{small}
\begin{equation}
\begin{aligned}
\label{eq-CapacityDiff-1}
    C_d =& C_{\hat{\mathcal{I}}} - C_{\mathcal{I}} \\
    =& N_r \log_2 \left( 1 + \frac{1}{N_0 \mathrm{tr}\left( (\mathbf{H}_{\hat{\mathcal{I}}} \mathbf{H}_{\hat{\mathcal{I}}}^H)^{-1} \right) } \right) \\
    &- N_r \log_2 \left( 1 + \frac{1}{N_0 \mathrm{tr}\left( (\mathbf{H}_{\mathcal{I}} \mathbf{H}_{\mathcal{I}}^H)^{-1} \right) } \right) \\
    =& N_r \log_2 \left( 1+ \frac{\mathrm{tr}\left( (\mathbf{H}_{\mathcal{I}} \mathbf{H}_{\mathcal{I}}^H)^{-1} \right) - \mathrm{tr}\left( (\mathbf{H}_{\hat{\mathcal{I}}} \mathbf{H}_{\hat{\mathcal{I}}}^H)^{-1} \right)}{\mathrm{tr}\left( (\mathbf{H}_{\hat{\mathcal{I}}} \mathbf{H}_{\hat{\mathcal{I}}}^H)^{-1} \right) \left(N_0 \mathrm{tr}\left( (\mathbf{H}_{\mathcal{I}} \mathbf{H}_{\mathcal{I}}^H)^{-1} \right) +1\right)} \right).
\end{aligned}    
\end{equation}%
\end{small}%
Based on block matrix multiplication, $\mathbf{H}_{\mathcal{I}} \mathbf{H}_{\mathcal{I}}^H = \mathbf{H}_{\hat{\mathcal{I}}} \mathbf{H}_{\hat{\mathcal{I}}}^H - \mathbf{H}_{\bar{\mathcal{I}}} \mathbf{H}_{\bar{\mathcal{I}}}^H$ holds, and by combining the SMW formula \eqref{eq-SMW}, it follows that 
\begin{equation}
\label{eq-SMW-HI}
(\mathbf{H}_{\mathcal{I}} \mathbf{H}_{\mathcal{I}}^H)^{-1} = (\mathbf{H}_{\hat{\mathcal{I}}} \mathbf{H}_{\hat{\mathcal{I}}}^H - \mathbf{H}_{\bar{\mathcal{I}}} \mathbf{H}_{\bar{\mathcal{I}}}^H)^{-1} = \mathbf{B} + \mathbf{D},
\end{equation}%
where $\mathbf{B} = (\mathbf{H}_{\hat{\mathcal{I}}} \mathbf{H}_{\hat{\mathcal{I}}}^H)^{-1}$, and $\mathbf{D} = \mathbf{B}\mathbf{H}_{\bar{\mathcal{I}}}(\mathbf{I} - \mathbf{H}_{\bar{\mathcal{I}}}^H \mathbf{B} \mathbf{H}_{\bar{\mathcal{I}}})^{-1} \mathbf{H}_{\bar{\mathcal{I}}}^H \mathbf{B}$. 
Substituting \eqref{eq-SMW-HI} into \eqref{eq-CapacityDiff-1}, we further obtain
\begin{equation}
\label{eq-CapacityDiff-2}
    C_d = N_r \log_2 \left( 1+ \frac{\mathrm{tr}\left( \mathbf{D} \right)}{\mathrm{tr}\left(\mathbf{B}\right) \mathrm{tr}\left((\mathbf{H}_{\mathcal{I}} \mathbf{H}_{\mathcal{I}}^H)^{-1}\right) N_0 + \mathrm{tr}\left(\mathbf{B}\right)} \right).   
\end{equation}%
Since $\mathbf{H}_{\mathcal{I}} \mathbf{H}_{\mathcal{I}}^H$ and $\mathbf{H}_{\hat{\mathcal{I}}} \mathbf{H}_{\hat{\mathcal{I}}}^H$ are both Hermitian matrices, $\mathrm{tr}\left((\mathbf{H}_{\mathcal{I}} \mathbf{H}_{\mathcal{I}}^H)^{-1}\right) > 0$ and $\mathrm{tr}\left(\mathbf{B}\right) >0$ always hold. In addition, according to the lemma in \cite{ref-TAS-PrecodedMIMO}, it is known that $\mathrm{tr}\left(\mathbf{D}\right) >0$ also always holds. 
Therefore, from \eqref{eq-CapacityDiff-2}, it can be observed that $C_d$ is monotonically decreasing with respect to the independent variable $N_0$. 
As the SNR tends to $\infty$, i.e., as $N_0 \to 0$, we can obtain the upper bound of the capacity loss as 
\begin{equation}
\label{eq-CapacityLoss}
    C_d < N_r \log_2 \left( 1+ \frac{\mathrm{tr}\left( \mathbf{D} \right)}{\mathrm{tr}\left(\mathbf{B}\right)} \right).   
\end{equation}%
Note that this upper bound is independent of the SNR. Therefore, given the index set of the activated ports, $\mathcal{I}$, the upper bound of the corresponding capacity loss can be calculated. As a result, \eqref{eq-CapacityLoss} can be used to guide the design of the FA-RSM system, helping to achieve a good trade-off between complexity and performance.

\subsection{Analysis of MMSE Precoding}
\label{Sec-PerformanceAnalysis-MMSE}
For the FA-RSM system with MMSE precoding, obtaining a closed-form solution for the capacity loss is difficult. Fortunately, the MMSE precoding matrix is designed by minimizing the MSE. Therefore, the MSE metric can replace capacity loss and be utilized to assist in evaluating the performance.

Given the index set of the activated ports $\mathcal{I}$, the MSE metric of MMSE-precoded systems has been provided in \cite{ref-MMSE-MSE} as
\begin{equation}
\begin{aligned}
\label{eq-MSE}
    \varepsilon_{\mathcal{I}} &= \mathrm{tr}\left( \left(\frac{1}{N_r N_0}\mathbf{H}_{\mathcal{I}} \mathbf{H}_{\mathcal{I}}^H + \mathbf{I}_{N_r}\right)^{-1} \right) \\
    &= N_r N_0 \mathrm{tr}\left( \left(\mathbf{H}_{\mathcal{I}} \mathbf{H}_{\mathcal{I}}^H + N_r N_0 \mathbf{I}_{N_r}\right)^{-1} \right).
\end{aligned}
\end{equation}%
As shown in \eqref{eq-MSE}, as the SNR $ \to \infty$, i.e., as $N_0 \to 0$, $\varepsilon_{\mathcal{I}}$ decreases and approaches 0.
Then, with the help of SMW formula \eqref{eq-SMW}, \eqref{eq-MSE} can be rewritten as
\begin{equation}
\begin{aligned}
\label{eq-MSE-SMW}
    \varepsilon_{\mathcal{I}} &= N_r N_0 \mathrm{tr}\left( \left(\mathbf{H}_{\hat{\mathcal{I}}} \mathbf{H}_{\hat{\mathcal{I}}}^H + N_r N_0 \mathbf{I}_{N_r}  - \mathbf{H}_{\bar{\mathcal{I}}} \mathbf{H}_{\bar{\mathcal{I}}}^H\right)^{-1} \right) \\
    &= N_r N_0 \left( \mathrm{tr}\left(\mathbf{B}^{\prime}\right) + \mathrm{tr}\left(\mathbf{D}^{\prime}\right) \right),
\end{aligned}    
\end{equation}%
where $\mathbf{B}^{\prime} = \left(\mathbf{H}_{\hat{\mathcal{I}}} \mathbf{H}_{\hat{\mathcal{I}}}^H + N_r N_0 \mathbf{I}_{N_r}\right)^{-1}$, and $\mathbf{D}^{\prime} = \mathbf{B}^{\prime}\mathbf{H}_{\bar{\mathcal{I}}}(\mathbf{I} - \mathbf{H}_{\bar{\mathcal{I}}}^H \mathbf{B}^{\prime} \mathbf{H}_{\bar{\mathcal{I}}})^{-1} \mathbf{H}_{\bar{\mathcal{I}}}^H \mathbf{B}^{\prime}$. 
As analyzed in Section \ref{Sec-PerformanceAnalysis-ZF}, similar to $\mathbf{D}$, $\mathbf{D}^{\prime}$ also satisfies $\mathrm{tr}\left(\mathbf{D}^{\prime}\right) >0$. 
Then, the MSE difference $\varepsilon_d$ between the port index sets $\hat{\mathcal{I}}$ and $\mathcal{I}$ can be given by
\begin{equation}
\begin{aligned}
\label{eq-MSEDiff}
    \varepsilon_d &= \varepsilon_{\mathcal{I}} - \varepsilon_{\hat{\mathcal{I}}} \\
    &= N_r N_0 \left( \mathrm{tr}\left(\mathbf{B}^{\prime}\right) + \mathrm{tr}\left(\mathbf{D}^{\prime}\right) \right) - N_r N_0 \mathrm{tr}\left(\mathbf{B}^{\prime}\right) \\
    &= N_r N_0 \mathrm{tr}\left(\mathbf{D}^{\prime}\right) \\
    &> 0.
\end{aligned}    
\end{equation}%
\eqref{eq-MSEDiff} demonstrates that a higher number of activated ports can reduce the MSE of the FA-RSM system. However, it can also be observed that as the SNR increases and tends to $\infty$, i.e., as $N_0 \to 0$, the MSE difference $\varepsilon_d$ decreases and approaches 0. This indicates that in high SNR scenarios, the gain from increasing the number of ports becomes very limited.

\section{Low-complexity Detector Design}  \label{Sec-DetectorDesign}
In this section, detection algorithms are proposed to reduce the complexity of the receiver in FA-RSM systems. 
First, the two-stage MED is introduced. 
Then, for MMSE precoding, the RTTD is further proposed to achieve a good trade-off between complexity and performance. 
The low-complexity detectors highlight the advantages of the RSM transmission mechanism.

\subsection{Two-stage MED}
As shown in \eqref{eq-MLD-ZF} and \eqref{eq-MLD-MMSE}, the complexity order of the search space for MLD is $\mathcal{O}(M N_r)$. 
Thus, to reduce the search complexity, by combining the characteristics of the FA-RSM received signal given in \eqref{eq-ReceivedSignal-ZF} and \eqref{eq-ReceivedSignal-MMSE}, we decouple the estimation of the receive antenna index $k$ and the demapping of the modulation symbol $s_m$, and propose the two-stage MED. 
Specifically, for the first stage of the MED, the detection result $\hat{k}$ of the receive antenna index is given by
\begin{equation}
\label{eq-MEDk}
    \hat{k} = \arg \underset{k}{\max} \left| y_k \right|^2.
\end{equation}%
For the second stage of the MED, by substituting the obtained $\hat{k}$ into \eqref{eq-MLD-ZF}, the demapping result for ZF precoding is obtained as
\begin{equation}
\label{eq-MEDsm-ZF}
    \hat{m} = \arg \underset{m}{\min} \left| y_{\hat{k}} - \beta_{\mathrm{ZF}} s_m \right|^2.
\end{equation}%
Substituting $\hat{k}$ into \eqref{eq-MLD-MMSE} gives the demapping result for MMSE precoding as
\begin{equation}
\label{eq-MEDsm-MMSE}
    \hat{m} = \arg \underset{m}{\min} \left| y_{\hat{k}} - \beta_{\mathrm{MMSE}} \mathbf{G}_{[\hat{k},\hat{k}]} s_m \right|^2.
\end{equation}%
Therefore, the MED reduces the complexity order of the search space to $\mathcal{O}(N_r + M)$.

\subsection{RTTD for MMSE Precoding}
When the FA-RSM system employs ZF precoding, the MED achieves performance that approximates MLD with lower complexity. 
However, when MMSE precoding is used, there is a significant performance gap between MED and MLD, as demonstrated in the subsequent simulation results. 
Therefore, we propose the RTTD as a tunable trade-off scheme for the FA-RSM system using MMSE precoding.

The energy of each component of the received vector $\mathbf{y}$ is ordered in decreasing sequence as $| y_{k_1} |^2 \geq | y_{k_2} |^2 \geq \ldots \geq | y_{k_{N_r}} |^2$, where $| y_{k_1} |^2$ and $| y_{k_2} |^2$ represent the components with the largest and second-largest received energies, respectively. 
The ratio of the second-largest value to the largest value is denoted as 
\begin{equation}
\label{eq-RTTD}
    r = | y_{k_2} |^2 / | y_{k_1} |^2,
\end{equation}%
with $r \in [0,1]$.
Then, $\gamma$ is used to denote the threshold of the ratio, with its value pre-determined by the Monte Carlo simulation results. 
If $r < \gamma$, then MED is adopted, with $\hat{k} = k_1$ and $\hat{m}$ given by \eqref{eq-MEDsm-MMSE}. 
Otherwise, the index information of the receive antenna with the maximum energy is considered to have low credibility, and in this case, MLD is adopted for detection, with $\hat{k}$ and $\hat{m}$ given by \eqref{eq-MLD-MMSE}. 

\section{Simulation Results}  \label{Sec-Simulation}

This section presents the simulation results of the proposed FA-RSM system, primarily including comparisons with the traditional RSM system, the effects and analysis of port selection, and the performance of the detectors.
Unless stated otherwise, the parameters of the FA-RSM system are set to the default values: $W_1 = W_2 = 1, N_1 = N_2 = 4, N_a = 4, N_r = 4$.
The FA-RSM system employs 4QAM by default, utilizes the optimal algorithm from Section \ref{Sec-PortSelect-Optimal} for port selection, and the optimal MLD described in Section \ref{SubSec-SystemModel-Detection} for detection. 

\subsection{Proposed FA-RSM Versus Traditional RSM}
\begin{figure}[t]
\centering
\subfloat[$N_a = 4$\label{fig-Result-VS_RSM_Na4}]{\includegraphics[width=0.9\linewidth]{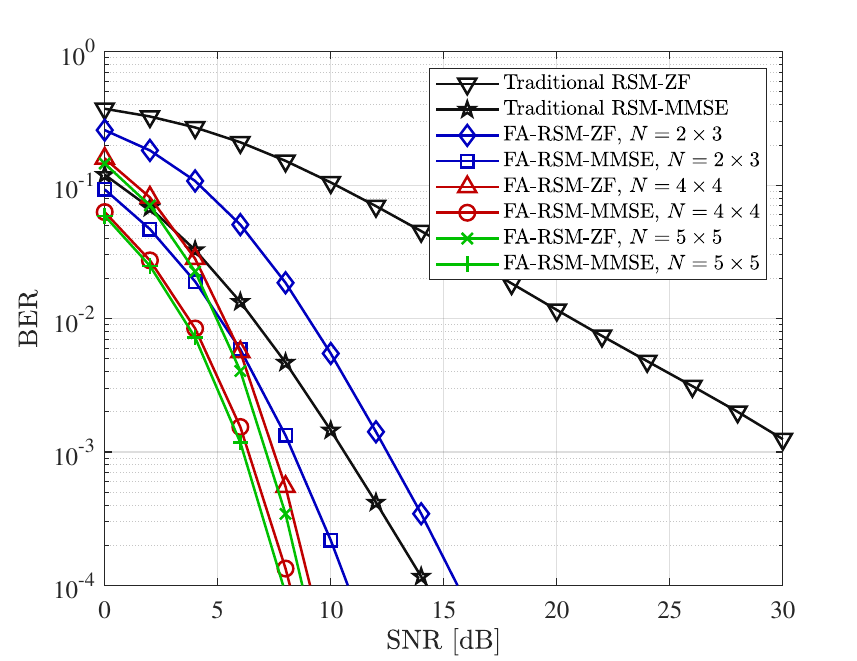}} \\
\subfloat[$N_a = 6$\label{fig-Result-VS_RSM_Na6}]{\includegraphics[width=0.9\linewidth]{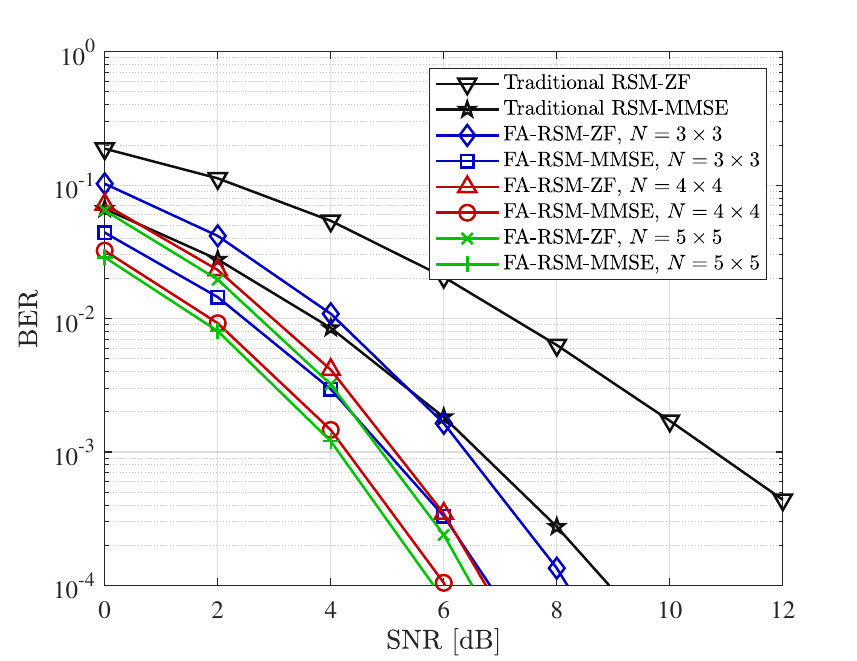}}
\caption{BER performance comparison between traditional RSM and proposed FA-RSM with varying port numbers $N$ for different $N_a$: (a) $N_a = 4$ and (b) $N_a = 6$.}
\label{fig-Result-VS_RSM}
\end{figure}
\figref{fig-Result-VS_RSM} compares the BER performance between the proposed FA-RSM and the traditional RSM. The RSM systems in \figref{fig-Result-VS_RSM_Na4} and \figref{fig-Result-VS_RSM_Na6} are equipped with 4 and 6 transmit antennas, respectively, while neglecting the spatial correlation between the antennas. FA-RSM-ZF and FA-RSM-MMSE are used to represent the FA-RSM systems with ZF precoding and MMSE precoding, respectively. Similarly, the curves for the RSM systems follow the same notation.
As shown in \figref{fig-Result-VS_RSM}, with the assistance of FA, the BER performance of the FA-RSM system significantly outperforms that of the RSM system. This is expected because the FA port selection provides additional spatial diversity. 
It is worth emphasizing that, compared to traditional RSM systems with no correlation, the FA-RSM systems in \figref{fig-Result-VS_RSM_Na4} with $N=6$ and in \figref{fig-Result-VS_RSM_Na6} with $N=9$ offer only 2 and 3 additional selectable ports, respectively, yet still deliver substantial gains.
Moreover, in both subfigures, it can be observed that distributing more ports within the given area enables the FA-RSM system to achieve higher performance gains. However, as $N$ increases, the performance gains achieved by adding more ports gradually diminish, which becomes particularly evident when comparing the curves for $N=16$ and $N=25$.
This phenomenon can be explained by the fact that, for a fixed $W$, increasing $N$ causes the ports to be placed closer together. The resulting mutual coupling effect leads to neighboring ports exhibiting similar channel characteristics. Therefore, blindly increasing the total number of ports is not always efficient.
Additionally, by comparing \figref{fig-Result-VS_RSM_Na4} and \figref{fig-Result-VS_RSM_Na6}, it is evident that increasing $N_a$ improves the performance of FA-RSM. However, due to the effects of spatial correlation, the performance gains of FA-RSM relative to RSM become smaller. This observation also motivates subsequent investigations into the impact of $N_a$ on the FA-RSM system.
In conclusion, \figref{fig-Result-VS_RSM} highlights the tremendous improvements that FA brings to the RSM system.

\subsection{Proposed Port Selection Algorithms}
\begin{figure}[t]
\centerline{\includegraphics[width=0.9\linewidth]{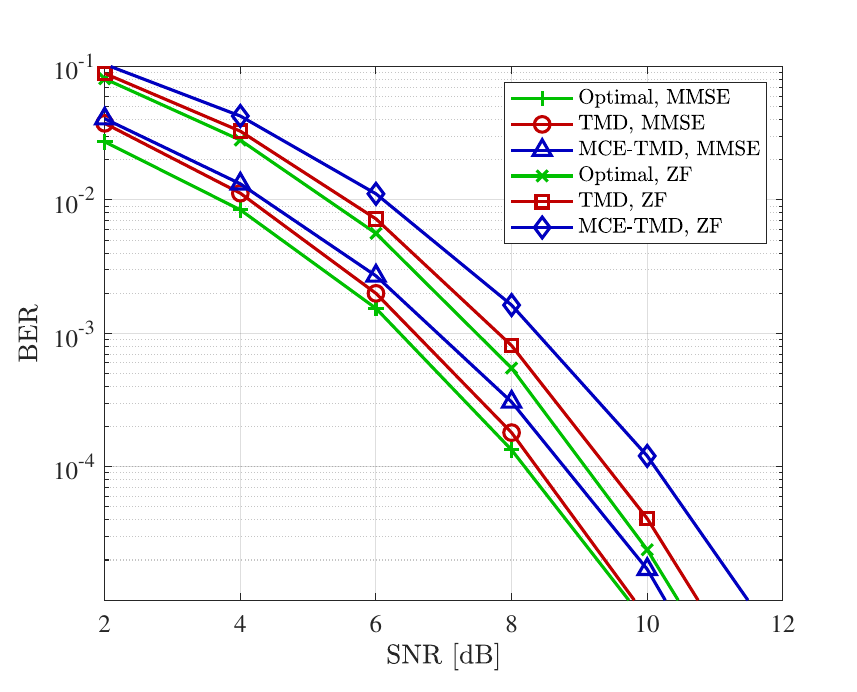}}
\caption{BER performance of the FA-RSM system with different port selection algorithms.}
\label{fig-Result-PortSelect}
\end{figure}
\figref{fig-Result-PortSelect} compares the impact of different port selection algorithms on the BER performance of the FA-RSM system, where $N_b=12$ in the MCE-TMD algorithm. 
It can be observed that, compared to the optimal algorithm, the performance loss of the two proposed low-complexity algorithms is relatively small. Additionally, the impact of the two algorithms on the performance of the system with MMSE precoding is smaller than their impact on the system with ZF precoding. 
Specifically, at a BER of $10^{-4}$, the performance degradation caused by the TMD algorithm for the MMSE and ZF systems is approximately 0.2 dB and 0.4 dB, respectively. The performance degradation caused by the MCE-TMD algorithm is approximately 0.6 dB and 1 dB, respectively. 
Therefore, both the TMD and MCE-TMD algorithms are efficient, offering excellent trade-offs between the performance and port selection complexity of the FA-RSM system.

\subsection{Impact of Port Selection}
\begin{figure}[t]
\centerline{\includegraphics[width=0.9\linewidth]{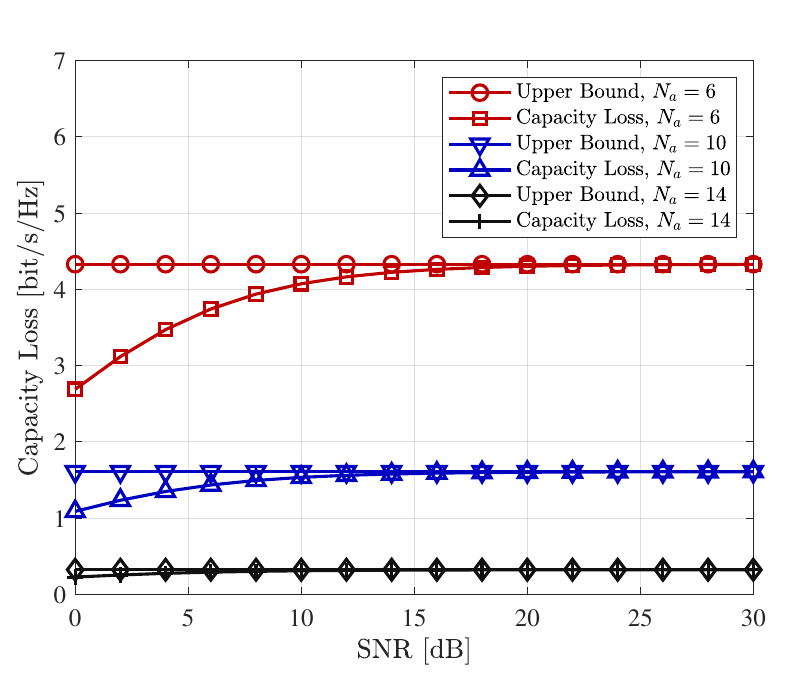}}
\caption{Capacity loss against SNR of FA-RSM with ZF precoding for different activated port numbers $N_a$.}
\label{fig-Result-CapLoss}
\end{figure}
\begin{figure}[t]
\centerline{\includegraphics[width=0.9\linewidth]{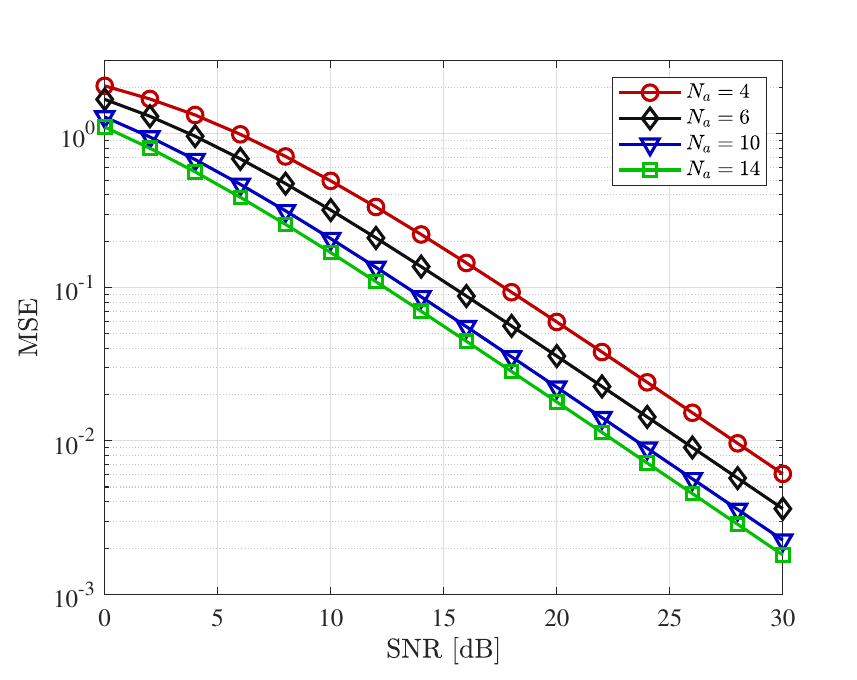}}
\caption{MSE against SNR of FA-RSM with MMSE precoding for different activated port numbers $N_a$.}
\label{fig-Result-MSE}
\end{figure}
\figref{fig-Result-CapLoss} and \figref{fig-Result-MSE} demonstrate the impact of port selection on the FA-RSM system from a theoretical perspective. 
Specifically, \figref{fig-Result-CapLoss} shows the capacity loss under ZF precoding and its derived upper bound, calculated using \eqref{eq-CapacityLoss}, for different numbers of activated ports $N_a$. 
As shown in \figref{fig-Result-CapLoss}, the capacity loss due to port selection is relatively small. As $N_a$ increases, the capacity loss approaches 0. Furthermore, in the high SNR region, all capacity losses reach their upper bounds and remain constant thereafter.
On the other hand, \figref{fig-Result-MSE} shows the MSE under MMSE precoding, calculated using \eqref{eq-MSE}, for different values of $N_a$.
In \figref{fig-Result-MSE}, the MSE monotonically decreases with increasing SNR. Additionally, it can be observed that as $N_a$ increases, the MSE decreases. However, the MSE improvement with increasing $N_a$ exhibits diminishing returns. 

\begin{figure}[t]
\centerline{\includegraphics[width=0.9\linewidth]{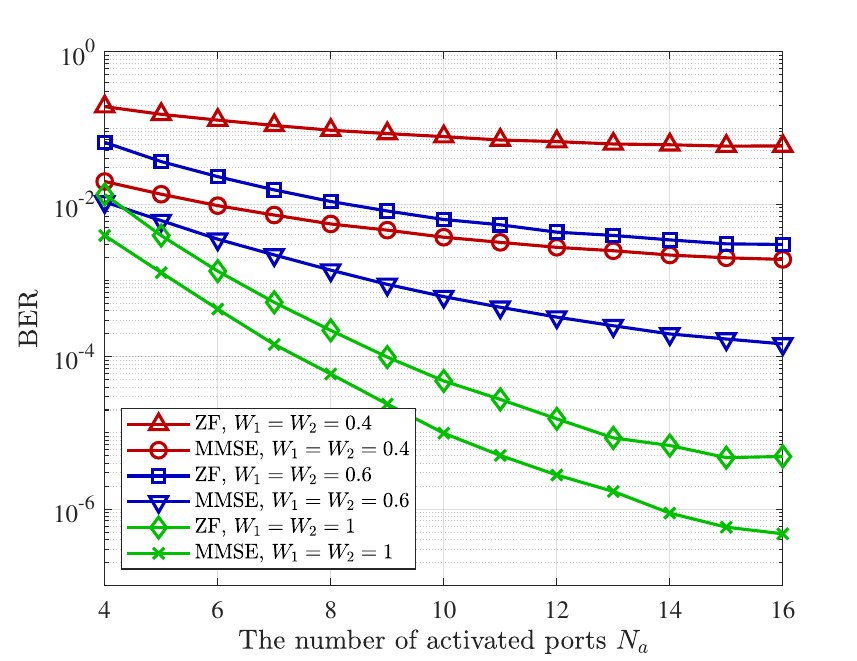}}
\caption{BER performance against activated port number for SNR = 5 dB and different $W_1$, $W_2$.}
\label{fig-Result-NaEffect}
\end{figure}
Next, \figref{fig-Result-NaEffect} evaluates the BER performance against the activated port number, $N_a$, under a fixed SNR of 5 dB. The simulation is conducted by altering the spatial correlation strength, i.e., keeping $N_1$ and $N_2$ constant and modifying $W_1$ and $W_2$.
It can be observed that, under a fixed SNR, increasing $N_a$ leads to a reduction in BER. However, the decrease in BER is not linear with respect to $N_a$; instead, a noticeable diminishing marginal effect is observed. This effect is more pronounced in the system using ZF precoding than in the system using MMSE precoding. Moreover, the effect becomes more significant as the spatial correlation strength increases, i.e., as $W_1$ and $W_2$ decrease.
Therefore, \figref{fig-Result-CapLoss}, \figref{fig-Result-MSE} and \figref{fig-Result-NaEffect} illustrate both theoretically and through simulations that activating more ports can improve the performance of the FA-RSM system. However, due to the high spatial correlation among FA ports, this improvement exhibits a significant diminishing marginal return. 
Consequently, it is crucial to carefully balance the cost, complexity, and performance of the FA-RSM system. The port selection algorithms and theoretical analysis presented in this paper provide effective tools for identifying a valuable trade-off.

\subsection{Proposed Detectors}
\begin{figure}[t]
\centering
\subfloat[SNR = -5 dB\label{fig-Result-gamma1}]{\includegraphics[width=0.9\linewidth]{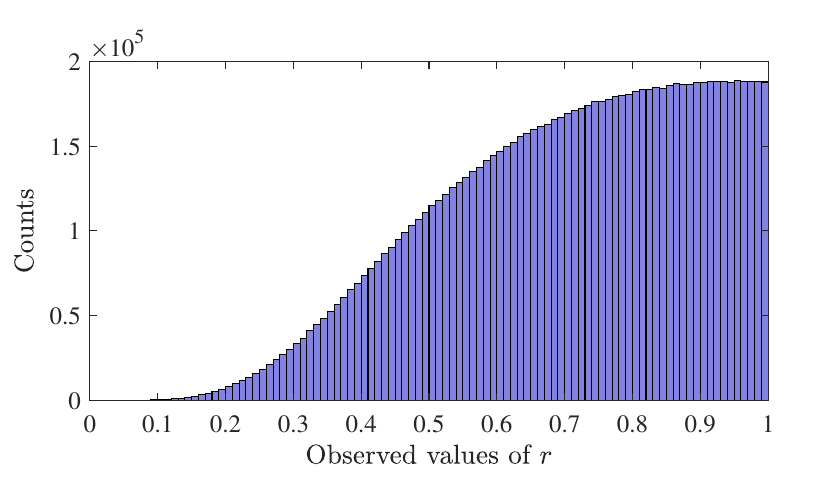}} \\
\subfloat[SNR = 10 dB\label{fig-Result-gamma2}]{\includegraphics[width=0.9\linewidth]{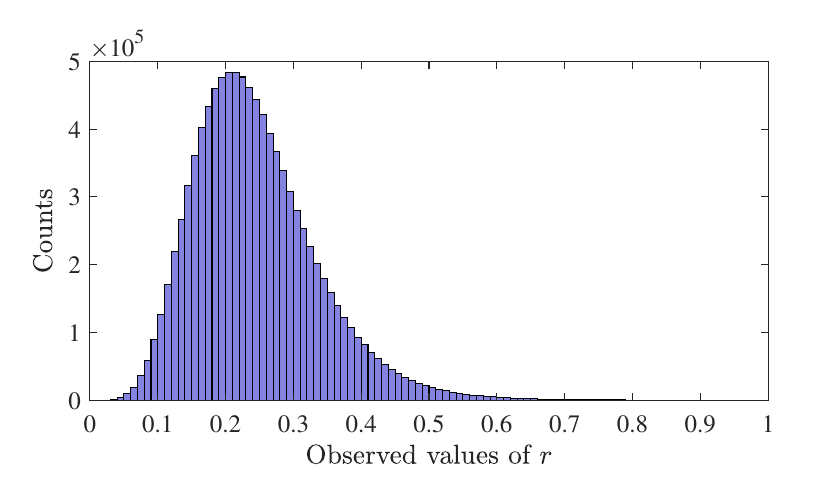}}
\caption{Histograms of the ratio $r$ for the FA-RSM system with MMSE precoding over $10^7$ trials: (a) SNR = -5 dB and (b) SNR = 10 dB.}
\label{fig-Result-gamma}
\end{figure}
Next, the performance of the detector will be evaluated. Before proceeding, it is essential to assign a value to the threshold parameter $\gamma$ in the proposed RTTD.
To this end, Monte Carlo simulations are performed over $10^7$ trials for the FA-RSM system with MMSE precoding at both relatively low and high SNRs. For each trial, the ratio $r$ of the second-largest to the maximum energy component of the received signal is computed using \eqref{eq-RTTD}, yielding the histograms shown in \figref{fig-Result-gamma}.
In \figref{fig-Result-gamma1}, the noise interference is relatively strong, causing the ratio $r$ to be skewed towards 1. In this scenario, the index information of the receive antenna with the maximum energy has low reliability in most trials, and thus, MLD should be employed more frequently to ensure reliable detection. 
In contrast, for \figref{fig-Result-gamma2}, where the noise interference is relatively weak, the distribution of $r$ is skewed towards 0. In this case, the index information of the receive antenna with the maximum energy is more reliable in most trials, making the MED a more suitable choice for achieving low-complexity detection.
Therefore, to strike a good balance between complexity and reliability, the threshold of RTTD is set to $\gamma=0.6$ for the FA-RSM system under the current configuration.

\begin{figure}[t]
\centerline{\includegraphics[width=0.9\linewidth]{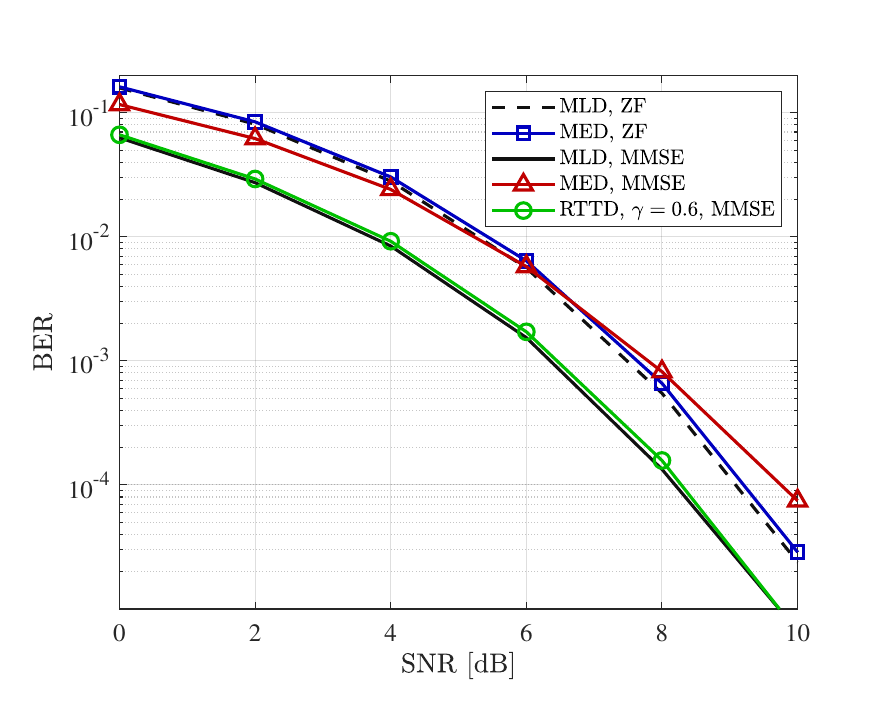}}
\caption{BER performance of the FA-RSM system with different detectors.}
\label{fig-Result-Detect}
\end{figure}
\figref{fig-Result-Detect} compares the BER performance of the proposed MED, RTTD, and the optimal MLD. 
As seen, for the FA-RSM system with ZF precoding, the use of the low-complexity MED achieves performance that closely approximates the optimal MLD. 
However, for the FA-RSM system with MMSE precoding, applying MED leads to significant performance degradation, which even results in worse performance  than the ZF precoded system using MED in high SNR regions. 
Fortunately, with the assistance of the RTTD designed for the MMSE precoded FA-RSM system, the system can achieve performance comparable to the optimal MLD at a lower complexity, effectively addressing the performance degradation caused by MED under MMSE precoding.
Therefore, the proposed MED and RTTD highlight the advantages of the FA-RSM system, where the receiver can achieve near-optimal detection performance with low complexity.

\section{Conclusion}  \label{Sec-Conclusion}
This paper proposed the FA-empowered RSM transmission system, where the transmitter is equipped with an FA, and the corresponding system model was developed. 
An optimal port selection algorithm was proposed from a capacity maximization perspective, along with two low-complexity alternatives: TMD and MCE-TMD. The latter further reduces complexity by exploiting the spatial correlation among ports. Compared to the optimal algorithm, both low-complexity algorithms incur minimal performance losses.
A theoretical analysis of port selection was provided under both ZF and MMSE precoding, supported by simulation results demonstrating that strong spatial correlation among FA ports makes port selection critical for balancing performance, cost, and complexity.
To reduce receiver complexity, the MED detector was proposed, and the RTTD detector was developed to mitigate the performance loss of MED under MMSE precoding.
These low-complexity detectors highlighted the receiver-friendly nature of the FA-RSM system. 
Future work may explore integrating the GSM principle to enhance SE, deploying FAs at both the transmitter and receiver, designing enhanced precoding schemes, and developing improved port selection algorithms and detectors under statistical CSI or imperfect channel estimation.




\bibliographystyle{IEEEtran}
\bibliography{IEEEabrv,mybib}

\end{document}